\documentclass{ws-ijmpe}
\usepackage[super,compress]{cite}
\usepackage[breaklinks]{hyperref}
\hypersetup{colorlinks,urlcolor=black,citecolor=black,linkcolor=black,filecolor=black}
\usepackage{breakurl}
\usepackage{url}
\usepackage[utf8]{inputenc}

\begin{document}

\markboth{R. Navarro P\'erez, J. E. Amaro and E. Ruiz Arriola}{Uncertainty quantification of effective nuclear interactions}

\catchline{}{}{}{}{}

\title{Uncertainty quantification of effective nuclear interactions}

\author{R. Navarro P\'erez}

\address{Nuclear and Chemical Science Division, Lawrence Livermore
  National Laboratory\\ Livermore, California 94551, USA
  \\ navarroperez1@llnl.gov}

\author{J. E. Amaro}

\address{Departamento de F\'{\i}sica At\'omica, Molecular y Nuclear
  and \\ Instituto Carlos I de F{\'\i}sica Te\'orica y Computacional,
  Universidad de Granada \\ E-18071 Granada, Spain \\ amaro@ugr.es}

\author{E. Ruiz Arriola}

\address{Departamento de F\'{\i}sica At\'omica, Molecular y Nuclear
  and \\ Instituto Carlos I de F{\'\i}sica Te\'orica y Computacional,
  Universidad de Granada \\ E-18071 Granada, Spain \\ earriola@ugr.es}

\maketitle

\begin{history}
\received{Day Month Year}
\revised{Day Month Year}
\end{history}

\begin{abstract}
We give a brief review on the development of phenomenological NN
interactions and the corresponding quantification of statistical
uncertainties. We look into the uncertainty of effective interactions
broadly used in mean field calculations through the Skyrme parameters
and effective field theory counter-terms by estimating both
statistical and systematic uncertainties stemming from the NN
interaction. We also comment on the role played by different fitting
strategies on the light of recent developments.  
\end{abstract}


\ccode{PACS numbers:03.65.Nk,11.10.Gh,13.75.Cs,21.30.Fe,21.45.+v}


\section{Introduction}

The study of nucleon-nucleon (NN) scattering acquired a central role
in nuclear physics with the first experimental measurements of
neutron-proton (np) and proton-proton (pp) differential cross sections
\cite{PhysRev.79.96, PhysRev.83.923}. Since then an ever increasing
database of NN scattering measurements at different kinematic
conditions has been collected in the literature and several
phenomenological potentials have been developed to describe it
\cite{Gammel:1957zz, Hamada:1962nq, Lassila:1962zz, Cottingham:1973wt,
  Machleidt:1987hj, Stoks:1994wp, Wiringa:1994wb, Machleidt:2000ge,
  Perez:2013jpa}. However, already in 1935 the seminal work of Yukawa
introduced the meson exchange picture where the NN interaction is the
result of the exchange of massive particles
\cite{yukawa1935interaction}. This is the basis of the well known one
pion exchange potential (OPE) which still nowadays gives the most
accurate description of the NN interaction at distances greater than
$3.0$ fm. The pioneering work of Gammel and Thaler in 1957 presented
an improvement over previous phenomenological approaches by including
a spin-orbit coupling term~\cite{Gammel:1957zz} and is considered the
first model with a semi-quantitative description of the
data~\cite{Machleidt:2001rw}. In later years several potentials,
including the ones from Hamada-Johnston~\cite{Hamada:1962nq},
Yale~\cite{Lassila:1962zz}, Paris~\cite{Cottingham:1973wt} and
Bonn~\cite{Machleidt:1987hj} presented gradual improvements by
including additional structural terms. For an in depth review of the
progress in NN phenomenological interactions before 1993 see
Refs.~\refcite{Machleidt:1989tm},~\refcite{Machleidt:1992uz} and
references therein. Despite the great theoretical efforts to obtain an
accurate representation of pp and np elastic scattering data, a
statistically successful description of such observables was not
possible until 1993 when the Nijmegen group discarded over a thousand
$3$ $\sigma$ inconsistent data out of 5246 np and pp scattering
observables and incorporated small but relevant electromagnetic
effects~\cite{Stoks:1993tb}. After this success a new generation of
realistic interactions were introduced including the NijmI, NijmII,
Reid93, ArgonneV18 and CD-Bonn potentials~\cite{Stoks:1994wp,
  Wiringa:1994wb, Machleidt:2000ge} as well as the covariant spectator
model~\cite{Gross:2008ps}. A least-squares merit figure yielding
$\chi^2/{\rm d.o.f.} \lesssim 1$ is a qualifying characteristic for a
potential to be considered realistic.  As opposed to phenomenological
interactions, a fundamental description of NN scattering data should
be given by the quantum chromodynamics (QCD) theoretical framework.
Using sub-nuclear degrees of freedom in terms of quarks and gluons it
is possible, in principle, to describe all levels of hadronic
interactions up to nuclear binding. However, despite the tremendous
effort done in direct lattice QCD calculations~\cite{Beane:2008ia,
  Aoki:2011ep, Aoki:2012tk} this approach still falls short when
confronted with NN scattering data (see however the recent
development~\cite{Berkowitz:2015eaa}). Other indirect approaches with
QCD ingredients like the inclusion of chiral symmetries in Effective
Field Theory (EFT)~\cite{Weinberg:1990rz, Ordonez:1993tn,
  Kaiser:1997mw, Rentmeester:1999vw, Entem:2003ft, Epelbaum:2004fk,
  Machleidt:2011zz} or large number of colors $N_c$ scaling for the NN
potential~\cite{Muther:1987sr, Kaplan:1996rk, Banerjee:2001js,
  CalleCordon:2008cz, CalleCordon:2009ps} either have a
phenomenological component or are unable to accurately reproduce NN
scattering observables as high quality interactions up to energies
near the pion production threshold~(see the discussions in
Refs.~\cite{Perez:2014bua,Piarulli:2014bda}).

Each experimental measurement of a NN scattering observable is subject
to random statistical fluctuations which are quantified by the
experimentalist in the form of error bars and in turn create a
statistical uncertainty in our knowledge of the NN interaction.  In
order to determine the size of this statistical uncertainty and
provide the NN potential with its corresponding error band an accurate
description of the over 8000 pp and np available published data is
necessary. Nowadays, phenomenological interactions represent the only
approach capable of such a description. Even though the study of the NN
interaction started more than six decades ago, the estimation of the
corresponding uncertainties has often been overlooked throughout all
those years. One of the main reasons behind the lack of estimation of
errors arising from experimental uncertainties in pp and np scattering
is probably the high level of complexity required to accurately
describe the NN interaction, especially when all the details of the
short-range or high momentum effects are taken into account
directly. Coarse graining embodies the Wilsonian
renormalization~\cite{Wilson:1973jj} concept and represents a very
reliable tool to simplify the description of pp and np scattering data
while still retaining all the relevant information of the interaction
up to a certain energy range set by the de Broglie wavelength of the
most energetic particle considered. The $V_{\mathrm{low} k}$
potentials in momentum space are a good example of an implementation
of coarse graining by removing the high-momentum part of the
interaction~\cite{Bogner:2003wn,Bogner:2009bt}.  Several potentials
and partial wave analyses (PWA) which accurately describe a large set
of pp and np scattering data can be found in the
literature~\cite{Rentmeester:1999vw, Ekstrom:2013kea, Entem:2003ft,
  Gross:2008ps, Machleidt:2000ge, Wiringa:1994wb, Stoks:1994wp,
  Stoks:1993tb}. Despite the great number of experimental data densely
probing the NN interaction in certain kinematic regions of energy and
scattering angle $(T_{\mathrm{LAB}},\theta)$ other areas of the same
plane remain mostly unexplored~\cite{Perez:2014yla}. This unbalance
creates an \emph{abundance bias} in which different phenomenological
potentials show agreement where experimental data are available but
great discrepancies can arise when predictions are made for the areas
where no data constrains the interaction. Also, each potential has its
own particular characteristics, some are given in momentum space,
others in coordinate space with different types of non-localities,
and some are energy dependent while others are not. These differences
in the theoretical representations of the NN interaction combined with
the abundance bias of the experimental data gives rise to significant
systematic uncertainties which propagate to any nuclear structure
calculation and therefore should be quantified to avoid performing
nuclear structure calculations with a superfluously high numerical
precision.

In this paper we will focus on a particular aspect of error analysis,
namely the determination of the low energy structure and its
statistical and systematic uncertainties from the point of view of
effective interactions.

\section{Effective Nuclear Interactions}
Power expansions in momentum space of effective interactions were
introduced by Moshinsky \cite{Moshinsky195819} and Skyrme
\cite{Skyrme1959615} to provide significant simplifications to the
nuclear many body problem in comparison with the {\it ab initio}
approach, in which it is customary to employ phenomenological
interactions fitted to NN scattering data to solve the nuclear many
body problem. As a consequence of such simplifications effective
interactions, also called Skyrme forces, have been extensively used in
mean field calculations \cite{Vautherin:1971aw, Negele:1972zp,
  Chabanat:1997qh, Bender:2003jk}. Within this framework the effective
force is deduced from the elementary NN interaction and encodes the
relevant physical properties in terms of a small set of
parameters. However, there is not a unique determination of the Skyrme
force and different fitting strategies result in different effective
potentials (see e.g. Refs.~\refcite{Friedrich:1986zza}
and~\refcite{Klupfel:2008af}). This diversity of effective
interactions within the various available schemes signals a source of
statistical and systematic uncertainties that remain to be
quantified. Fortunately the parameters determining a Skyrme force can
be extracted from phenomenological interactions \cite{Arriola:2010hj,
  NavarroPerez:2013iwa} and uncertainties can be propagated
accordingly \cite{Perez:2014kpa}. At the two body level the
Moshinsky-Skyrme potential in momentum representation reads
\begin{eqnarray} 
 V_\Lambda ({\bf
    p}',{\bf p}) 
&=& \int d^3 x e^{-i {\bf x}\cdot ({\bf p'}-{\bf p})}  \hat V({\bf x} ) 
 \nonumber \\ &=&  t_0 (1 + x_0 P_\sigma ) + \frac{t_1}2(1 + x_1
  P_\sigma ) ({\bf p}'^2 + {\bf p}^2) \nonumber \\ 
&&+  
 t_2 (1 + x_2
  P_\sigma ) {\bf p}' \cdot {\bf p} + 2 i W_0 {\bf S} \cdot({\bf p}'
  \times {\bf p}) \nonumber \\ &&+ 
\frac{t_T}2 \left[ \sigma_1 \cdot {\bf p}
  \, \sigma_2 \cdot {\bf p}+ \sigma_1 \cdot {\bf p'} \, \sigma_2
  \cdot {\bf p'} - \frac13 \sigma_1 \, \cdot 
\sigma_2 ({\bf p'}^2+  {\bf p}^2)
\right] \nonumber \\  &&+
\frac{t_U}2 \left[ \sigma_1 \cdot {\bf p}
  \, \sigma_2 \cdot {\bf p}'+ \sigma_1 \cdot {\bf p'} \, \sigma_2
  \cdot {\bf p} - \frac23 \sigma_1 \, \cdot 
\sigma_2 {\bf p'}\cdot  {\bf p}
\right]  
+ {\cal O} (p^4) 
\label{eq:skyrme2}
\end{eqnarray} 
where $P_\sigma = (1+ \sigma_1 \cdot \sigma_2)/2$ is the spin exchange
operator with $P_\sigma=-1$ for spin singlet $S=0$ and $P_\sigma=1$
for spin triplet $S=1$ states. The cut-off $\Lambda$ specifies the
maximal CM momentum scale, and therefore $\Delta r=\hbar/\Lambda$
determines the de Broglie resolution.

As mentioned above different nuclear data can be used to constrain the
Skyrme potential. The usual approach is to fit parameters of
Eq.~(\ref{eq:skyrme2}) to doubly closed shell nuclei and nuclear
matter saturation properties~\cite{Vautherin:1971aw, Negele:1972zp,
  Chabanat:1997qh, Bender:2003jk}. In Ref.~\refcite{Arriola:2010hj}
the parameters were determined from just NN threshold properties such
as scattering lengths, effective ranges and volumes without explicitly
taking into account the finite range of the NN interaction; while in
Ref.~\refcite{NavarroPerez:2013iwa} the parameters were computed
directly from a local interaction in coordinate space that reproduces
NN elastic scattering data. In Ref.~\refcite{Perez:2014kpa} the latter
approach was used to propagate \emph{statistical} uncertainties into
the Skyrme parameters. Here we will follow along the same technique to
quantify the \emph{systematic} uncertainties, which arise from the
different representations of the NN interaction. For completeness we
reproduce the equations necessary to compute the Skyrme parameters
from a local potential in coordinate space:
\begin{eqnarray}
(t_0, x_0 t_0) &=& \frac12 \int\, d^3 x\,  \,  \left[ V_{^3S_1}(r) \pm V_{^1S_0}(r) \right] 
\, , \nonumber \\  
(t_1, x_1 t_1)  &=& -\frac1{12} \int\, d^3 x\, r^2 \,  \left[ V_{^3S_1}(r) \pm V_{^1S_0}(r) \right] 
\, , \nonumber \\  
(t_2, x_2 t_2)  &=& \frac{1}{54} \int \, d^3 x\, r^2 \, \left[
  V_{^3P_0}(r) + 3 V_{^3P_1}(r) + 5 V_{^3P_2}(r)\pm 9 V_{^1P_1}(r)
  \right] \, , \nonumber \\   
 W_0 &=& \frac{1}{72} \int \, d^3 x\,  r^2 \, \left[ 2 V_{^3P_0} (r) +  3 V_{^3P_1} (r) -5  V_{^3P_2} (r) \right] \, , \nonumber \\  
t_T &=& \frac{1}{5 \sqrt{2}} \int \, d^3 x\, r^2 \,  V_{E_1}(r) \, , \nonumber \\t_U &=& \frac{1}{36} \int \, d^3 x\, r^2 \, \left[
-2  V_{^3P_0}(r) + 3 V_{^3P_1}(r) - V_{^3P_2}(r)\right] \, , 
\label{eq:skyrme}
\end{eqnarray}
where the $\pm$ in the first three equations refers to the first
and second possibilities on the l.h.s.

Alternatively we consider effective interactions derived from a low
momentum interaction where the coefficients can be identified with the
phenomenological counter-terms of chiral effective field theory. To
obtain such counter-terms we express the momentum space NN potential
in the partial wave basis
\begin{eqnarray}
v^{JS}_{l',l} (p',p) =(4\pi)^2 \int_0^\infty \, dr\, r^2 \, j_{l'}(p'r)
j_{l}(pr) V_{l' l}^{JS}(r) \,
\end{eqnarray}
and use the Taylor expansion of the spherical Bessel function 
\begin{eqnarray}
j_l(x) = \frac{x^l}{(2l+1)!!}  \left[1 - \frac{x^2}{2(2l+3)} + \dots
  \right]
\end{eqnarray}
to get an expansion for the potential in each partial wave. Keeping
terms up to fourth order ${\cal O} (p^4, p'^4, p^3p', pp'^3, p^2p'^2)$
corresponds to keeping only $S$-, $P$- and $D$-waves along with
$S$-$D$ and $P$-$F$ mixing parameters.  Using the normalization and
spectroscopic notation of Ref.~\refcite{Epelbaum:2004fk} one gets
\begin{eqnarray}
v_{00}^{JS}(p',p) 
&=& 
 \widetilde{C}_{00}^{JS} 
+ C_{00}^{JS}(p^2+p'^2) 
+ D^1_{00}{}^{JS} (p^4+p'^4) 
+ D^2_{00}{}^{JS} p^2 p'^ 2 
+ \cdots 
\nonumber \\ 
v_{11}^{JS}(p',p) 
&=& 
p p' C_{11}^{JS} 
+ p p' (p^2+p'^2) D_{11}^{JS} 
+ \cdots 
\nonumber \\
v_{22}^{JS}(p',p) 
&=& 
p^2 p'{}^2 D_{22}^{JS} 
+ \cdots 
\nonumber   \\ 
v_{20}^{JS}(p',p) 
&=& p'{}^2 C_{20}^{JS} 
+ p'{}^2 p^ 2  D^1_{20}{}^{JS} + p'{}^4 D^2_{20}{}^{JS} 
+ \dots \nonumber   \\
v_{31}^{JS}(p',p) 
&=& 
p'{}^3 p D_{31}^{JS} 
+ \cdots 
\label{eq:Cts}
\end{eqnarray}
and each counter-term can be expressed as a radial momentum of the NN
potential in a specific partial wave. Different methods have been
proposed to quantify some of the uncertainties in these
quantities\cite{Epelbaum:2014efa, Furnstahl:2015rha}. In this work we
follow a direct procedure that completely determines the relevant
uncertainties.

\subsection{Statistical Uncertainties}

Most phenomenological interactions are determined by a least squares
procedure consisting of the minimization of the figure of merit
\begin{equation}
\chi^2 = \sum_{i=1}^N R_i^2 = \sum_{i=1}^N \frac{(O_i^{\rm exp} -
  O_i^{\rm theor}(p_1,p_2,\ldots,p_P))^2}{(\Delta O_i^{\rm exp})^2},
\label{eq:chi2-exp}
\end{equation}
where $O_i^{\rm exp}$ are the measured data with an experimental error
$\Delta O_i^{\rm exp}$, $O_i^{\rm theor}$ are the theoretical values
determined by the potential parameters $\pmb{p} =
(p_1,p_2,\ldots,p_P)$ and $R_i$ are known as residuals. The
minimization of $\chi^2$ corresponds to finding the most likely values
for the fitting parameters $\pmb{p}_0$ given by
\begin{equation}
 \min \chi^2 (\pmb{p}) = \min \sum_{i=1}^N
 \frac{(O_i^{\rm exp} - O_i^{\rm theor}(\pmb{p}))^2}{(\Delta O_i^{\rm
     exp})^2} = \chi^2 (\pmb{p}_0).
\end{equation}
In practice an agreement between theory and experiment requires
$\chi^2/(N-P) \sim 1$.  But this does not quantify the uncertainty of
the parameters after the fit.  To obtain statistically justified
uncertainties the following crucial property must hold: The
discrepancies between theoretical and experimental values are
independent and normally distributed. This assumption of course can
only be checked after the possibly complex and numerically
expensive process of fitting the interaction to reproduce
experimental NN scattering data. However, testing for the normality of
the residuals is an easy and straightforward procedure, as detailed in
Ref.~\refcite{Perez:2014kpa}, with a wide range of different tests
available in the literature and several of them already implemented in
mathematical packages.

Once the assumption of normally distributed residuals has been
positively tested it is possible to propagate the experimental
uncertainty into the potential parameters in the form of confidence
regions via the standard procedure of obtaining the parameters'
covariance matrix\footnote{If the residuals are shown to not follow
  the standard normal distribution different error propagation
  techniques have to be used. See for example the Bayesian method
  detailed in Ref.~\refcite{Furnstahl:2014xsa} and employed recently
  in Ref.~\refcite{Furnstahl:2015rha} }. These confidence regions
contain the set of values $\pmb{p}'$ around $\pmb{p}_0$ that give
$\chi^2(\pmb{p}') \leq \chi^2(\pmb{p}_0) + 1$ and represent the
statistical uncertainty of the phenomenological interaction. The
covariance matrix is defined as the inverse of the Hessian matrix
\begin{equation}
  ({\cal C}^{-1})_{i j} \equiv H_{i j} = \frac{\partial
    \chi^2}{\partial p_i \partial p_j}
\end{equation}
and can be used to calculate the statistical uncertainty of any quantity
expressed as a function of the potential parameters
\begin{equation}
 (\Delta F)^2 = \sum_{i j} \frac{\partial F}{\partial
    p_i}\frac{\partial F}{\partial p_j} {\cal C}_{i j,}.
  \label{eq:CovarianceError}
\end{equation}

Our series of phenomenological interactions, including the DS-OPE,
DS-$\chi$TPE, DS-Born, Gauss-OPE, Gauss-$\chi$TPE and
Gauss-Born\cite{Perez:2013jpa, Perez:2013mwa, Perez:2013oba,
  Perez:2014yla, Perez:2014waa}, have all been positively and
stringently tested for normally distributed residuals. This allows us to
confidently propagate the statistical uncertainties into the Skyrme
parameters of Eq.(\ref{eq:skyrme}) and the counter-terms of
Eq.(\ref{eq:Cts}) using the parameters' covariance matrix as indicated
in Eq.(\ref{eq:CovarianceError}). Our results are summarized in
tables~\ref{tab:Skyrme} and~\ref{tab:Cterms}. The statistical
uncertainties in both the Skyrme parameters and the counter-terms are
about the same order of magnitude for the six potentials
considered. This was expected since all six interactions are
statistically equivalent in the sense that each one describes the
self-consistent database with $\chi^2/(N-P) \sim 1$ and their
residuals follow the standard normal distribution.

\begin{table}[pt]
\tbl{Moshinsky-Skyrme parameters for the renormalization scale
  $\Lambda=400$ MeV. Errors quoted for each potential are statistical;
  errors in the last column are systematic and correspond to the
  sample standard deviation of the six previous columns. See main text
  for details on the calculation of systematic errors. Units are:
  $t_0$ in ${\rm MeV} {\rm fm}^3$, $t_1,t_2,W_0,t_U,t_T$ in ${\rm MeV}
  {\rm fm}^5$, and $x_0,x_1,x_2$ are
  dimensionless. \label{tab:Skyrme}}
    {\begin{tabular}{@{}cccccccc@{}}
    \toprule & DS-OPE & DS-$\chi$TPE& DS-Born & Gauss-OPE
    &Gauss-$\chi$TPE& Gauss-Born & Compilation \\ \colrule $t_0$ &
    -626.8(64) & -529.6(53) & -509.0(55) & -584.4(157) & -406.1(289) &
    -521.8(152) & -529.6(751) \\ $x_0$ & -0.38(2) & -0.56(1) &
    -0.54(1) & -0.26(2) & -0.71(8) & -0.55(4) & -0.50(16) \\ $t_1$ &
    948.1(30) & 913.6(22) & 900.1(17) & 987.4(29) & 945.5(18) &
    941.3(16) & 939.3(304) \\ $x_1$ & -0.048(3) & -0.074(3) &
    -0.068(3) & -0.013(3) & -0.047(3) & -0.058(2) & -0.051(22)\\ $t_2$
    & 2462.6(56) & 2490.0(39) & 2462.1(25) & 2441.3(56) & 2490.1(24) &
    2466.8(26) & 2468.8(187) \\ $x_2$ & -0.8686(6)& -0.8750(8)&
    -0.8753(6)& -0.8630(8)& -0.8729(6) & -0.8785(3)& -0.872(6)
    \\ $W_0$ & 107.7(4) & 100.8(3) & 96.2(3) & 105.0(5) & 109.3(7) &
    94.3(2) & 102.2(61) \\ $t_U$ & 1278.6(12) & 1260.3(5) & 1257.0(4)
    & 1285.6(12) & 1254.9(9) & 1249.3(3) & 1264.3(144) \\ $t_T$
    &-4220.9(87) &-4292.8(23) &-4289.0(21) &-4385.6(99) & -4271.8(51)
    &-4319.5(58) & -4296.6(545) \\ \botrule
\end{tabular}}
\end{table}

\begin{table}[pt]
\tbl{Potential integrals in different partial waves. Errors quoted for
  each potential are statistical; errors in the last column are
  systematic and correspond to the sample standard deviation of the
  six previous columns. See main text for details on the calculation
  of systematic errors. Units are: $\widetilde{C}$'s are in $10^4$
  ${\rm GeV}^{-2}$, $C$'s are in $10^4$ ${\rm GeV}^{-4}$ and $D$'s are
  in $10^4$ ${\rm GeV}^{-6}$. \label{tab:Cterms}}
    {\begin{tabular}{@{}cccccccc@{}} \toprule &DqS-OPE &DS-$\chi$TPE&
        DS-Born & Gauss-OPE &Gauss-$\chi$TPE&Gauss-Born &
        Compilation\\ \colrule $\widetilde{C}_{^1S_0}$ & -0.141(1)&
        -0.135(2) & -0.128(2)& -0.121(5)& -0.113(9) & -0.133(3)&
        -0.13(1) \\ $C_{^1S_0}$ & 4.17(2) & 4.12(2) & 4.04(1) &
        4.20(2) & 4.16(2) & 4.18(1) & 4.15(6) \\ $D_{^1S_0}^1$
        &-448.8(11) & 443.7(5) &-441.5(3) &-447.0(10) & -446.7(2)
        &-446.3(2) &-445.7(26) \\ $D_{^1S_0}^2$ &-134.6(3) &-133.1(1)
        &-132.46(4) &-134.1(3) & -134.02(7) &-133.90(7) &-133.7(8)
        \\ $\widetilde{C}_{^3S_1}$ & -0.064(2)& -0.038(1) & -0.039(1)&
        -0.070(2)& -0.019(6) & -0.038(4)& -0.045(19)\\ $C_{^3S_1}$ &
        3.79(1) & 3.55(1) & 3.52(1) & 4.09(2) & 3.785(9) & 3.724(9)&
        3.7(2) \\ $D_{^3S_1}^1$ &-510.7(3) &-504.7(4) &-504.1(2)
        &-516.7(6) & -509.7(1) &-508.2(1) &-509.0(46) \\ $D_{^3S_1}^2$
        &-153.2(1) &-151.4(1) &-151.22(6) &-155.0(2) & -152.90(3)
        &-152.47(3) &-152.7(14) \\ $C_{^1P_1}$ & 6.44(2) & 6.54(1) &
        6.464(6)& 6.37(2) & 6.529(7) & 6.488(7)& 6.47(6)
        \\ $D_{^1P_1}$ &-594.9(2) &-592.1(2) &-590.21(6) &-594.5(2) &
        -597.83(7) &-596.25(7) &-594.3(28) \\ $C_{^3P_1}$ & 3.738(2)&
        3.659(3) & 3.633(3)& 3.762(6)& 3.677(3) & 3.599(1)& 3.68(6)
        \\ $D_{^3P_1}$ &-253.29(5) &-249.8(2) &-249.62(7) &-254.23(9)
        & -251.0(2) &-251.06(2) &-251.5(19) \\ $C_{^3P_0}$ &
        -4.911(8)& -4.882(5) & -4.897(3)& -4.944(6)& -4.802(8) &
        -4.883(2)& -4.89(5) \\ $D_{^3P_0}$ & 347.0(2) & 343.6(2) &
        344.62(6) & 345.8(1) & 345.02(3) & 346.25(2) & 345.4(12)
        \\ $C_{^3P_2}$ & -0.445(2)& -0.434(3) & -0.426(2)& -0.426(2)&
        -0.448(1) & -0.427(1)& -0.43(1) \\ $D_{^3P_2}$ & -10.62(7) &
        -9.7(2) & -9.45(6) & -11.55(4) & -9.939(8) & -9.631(7)&
        -10.1(8) \\ $D_{^1D_2}$ & -70.92(3) & -70.66(6) & -70.52(3) &
        -70.58(3) & -71.109(7) & -71.074(5)& -70.8(3) \\ $D_{^3D_2}$
        &-367.8(2) &-364.39(7) &-364.54(4) &-367.19(8) & -367.10(2)
        &-366.99(1) &-366.3(15) \\ $D_{^3D_1}$ & 205.8(2) & 204.25(7)
        & 204.26(4) & 204.4(1) & 205.17(3) & 205.21(3) & 204.9(6)
        \\ $D_{^3D_3}$ & 0.55(1) & 0.87(6) & 0.90(4) & -0.32(9) &
        0.26(3) & 0.51(3) & 0.46(45) \\ $C_{\epsilon_1}$ & -8.36(2) &
        -8.500(4) & -8.492(4)& -8.35(1) & -8.404(4) & -8.399(5)&
        -8.42(7) \\ $D_{\epsilon_1}^1$ &1012.6(6) &1005.5(1)
        &1006.23(6) &1010.5(3) & 1011.83(5) &1012.71(6) &1009.9(32)
        \\ $D_{\epsilon_1}^2$ & 434.0(3) & 430.94(4) & 431.24(3) &
        433.1(1) & 433.64(2) & 434.02(2) & 432.8(14)
        \\ $D_{\epsilon_2}$ & 84.18(4) & 83.29(1) & 83.398(7)&
        84.25(3) & 83.660(5) & 83.818(8)& 83.8(4) \\ \botrule
\end{tabular}}
\end{table}
 
\subsection{Systematic Uncertainties}

\begin{figure}[th]
\centerline{\includegraphics[width=10cm]{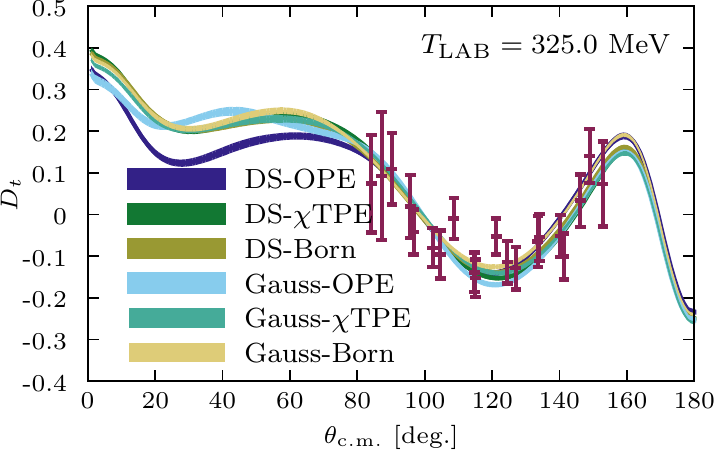}}
\caption{Polarization transfer parameter $D_t$ at $T_{\rm LAB} = 325$
  MeV as a function of center of mass scattering angle $\theta_{\rm
    c.m.}$. The six bands correspond to the predictions of the six
  statistically equivalent phenomenological potentials DS-OPE (dark
  blue), DS-$\chi$TPE (dark green), DS-Born (dark yellow), Gauss-OPE
  (light blue), Gauss-$\chi$TPE (light green) and Gauss-Born (light
  yellow). The predictions are compared with experimental
  data (red error bars)~\cite{Clough:1980hs, Amsler:1977nx}.
 \label{fig:DtObservable}}
\end{figure}

As mentioned earlier the uneven distribution of experimental NN
scattering data over the laboratory energy - scattering angle plane
creates an abundance bias in which the discrepancies of predictions by
phenomenological interactions for observables in unexplored kinematic
regions can be significantly larger than the inferred statistical
uncertainties.  A clear example of this abundance bias is shown in
Fig.~\ref{fig:DtObservable} which plots the polarization transfer
parameter $D_t$ at $T_{\rm LAB} = 325$ MeV as a function of the center
of mass scattering angle $\theta_{\rm c.m.}$.  For forward scattering
angle, where no measurements are available, the six phenomenological
potentials show incompatible predictions. This signals the presence of
systematic uncertainties arising from the different representations of
the NN interaction. In the particular case of the phenomenological
interactions listed in tables~\ref{tab:Skyrme} and~\ref{tab:Cterms}
the six potentials can be considered to be statistically equivalent
since each one gives an accurate description of the \emph{same}
self-consistent database with normally distributed
residuals. Therefore any discrepancy in their predictions can only be
attributed to the different representations of the short, intermediate
and long range parts. These discrepancies have been observed in
phase-shifts, scattering amplitudes and low energy parameters to be
considerably larger than the corresponding statistical
uncertainties\cite{Perez:2014waa}.

To estimate the systematic uncertainty on the Skyrme parameters and
counter-terms of six statistically equivalent phenomenological
interactions presented in the previous section we use the sample 
standard deviation defined as
\begin{equation}
s = \sqrt{\frac{1}{N-1}\sum_{i=1}^N(x_i-\bar{x})^2},
\end{equation}
where $\bar{x}$ is the usual sample mean. The reason for the $1/(N-1)$
factor, instead of the usual $1/N$, is to reduce the bias generated
from having an estimate of the mean from a sample of six interactions
instead of the actual mean of the population. Of course, this
procedure can only give a lower limit of the actual systematic
uncertainty since only a subset of \emph{all} the possible
statistically equivalent representations of the NN interaction can be
included. Our results are presented in the last column of
tables~\ref{tab:Skyrme} and~\ref{tab:Cterms}, with the sample mean as
central value. The estimated systematic uncertainties for the Skyrme
parameters and counter-terms are always at least an order of magnitude
larger than the statistical ones from each one of the six potentials
considered. This is in agreement with the results of
Ref.~\refcite{Perez:2014waa} in phase-shifts, scattering amplitudes
and low energy parameters.

\section{Discussion on fitting strategies}

Our estimates on the statistical and systematic errors of both Skyrme
parameters and counter-terms are based on 6 different, statistically
significant and equivalent fits to a set of $\sim 6700 $ np and pp
scattering data below lab energies of $350$ MeV, the canonical and
traditional upper energy for NN potentials marked by pion production
threshold. They are representative of similar features found
previously~\cite{Perez:2014waa} for phase-shifts, scattering
amplitudes and low energy threshold parameters.

\subsection{Consistent vs inconsistent fits}

Since the earliest fits, the validation of the NN interaction has
proceeded normally by the well known least squares $\chi^2$-method.
This is certainly a convenient and handy way of minimizing
discrepancies between the theoretical model characterized by unknown
parameters and the available experimental data with their assumed
uncertainties. One of the great advantages of the approach is that it
has a probabilistic interpretation and relies on a maximum likelihood
principle. In the observable space $(O_1, \dots O_N)$ the $\chi^2$
could be interpreted as a distance enjoying all necessary axioms of a
metric measuring the separation between a given theory $O_i^{\rm
  theor}(p_1,p_2,\ldots,p_P)$ and the experimental result $O_i^{\rm
  exp}$ in units of the experimental uncertainties $\Delta O_i^{\rm
  exp}$. Therefore we may understand a sense of proximity within the
space of parameters $(p_1, p_2,\ldots,p_P)$. Precisely because of that
this is not the only sense of proximity possible, and any other least
distance principle might be used to optimize the theory~\footnote{For
  instance one may choose to minimize just the absolute value of the
  difference or take some convenient power $q$
\begin{eqnarray}
d_q (O^{\rm exp}, O^{\rm th})= \left\{ \sum_{i=1}^N \Big|\frac{O_i^{\rm exp} -
  O_i^{\rm theor}(p_1,p_2,\ldots,p_P)}{\Delta O_i^{\rm exp}} \Big|^q \right\}^{\frac{1}q}
\end{eqnarray}
with the purpose of, say, excluding outliers, or serving some other
purpose.}. Of course, alternative approaches provide different senses
of optimization but also require more detailed studies and often do
not go beyond useful recipes which can hardly be justified {\it a
  priori} nor checked {\it a posteriori}.

In contrast, the popularity enjoyed by the traditional least squares
approach relies actually in this capability of checking the
self-consistency. This is admittedly a potential drawback for the
utility of the approach, since one may end up in the uncomfortable
situation of having a visually good fit but a mathematically
inconsistent one. The fits carried out by ourselves also run these
risks. However, we wonder whether it makes sense to take just the
advantages of the $\chi^2$ without buying the disadvantages. This
remark applies beyond the NN data analysis.~\footnote{ In the simple
  case of $\pi\pi$ scattering the many possible pitfalls of such an
  attitude have been illustrated~\cite{Perez:2015pea}.} As a
consequence of this, many of the visually good but inconsistent fits
that were found in the process of the present and previous
investigations have not been reported.  For instance a fit to the full
database, without the $3$ $\sigma$ rejection criterion invoked
previously~\cite{Perez:2013jpa}, provides a $\chi^2/{\rm d.o.f.} =1.7$
for about $N \sim 8000$ NN scattering data and corresponds to a
discrepancy with the chi square distribution at the $25$ $\sigma$
level. As shown later the re-scaling by a Birge
factor~\cite{PhysRev.40.207} {\it cannot} be applied as the normality
test fails.

Some of the renowned and benchmarking papers on NN fits do not even
quote their best $\chi^2$ value~\cite{Hamada:1962nq,Machleidt:1987hj},
and it is unclear what they actually mean by improvement beyond a
purely visual and subjective inspection of the plots.

For instance, the SAID analysis~\cite{SAID} yields $\chi^2/N \sim 1.7$
for the analyzed data bellow laboratory energy of $350$ MeV,
which is $25$ $\sigma$ away from any reasonable confidence level. We do
not exclude in principle the possibility that a normality test
analysis on residuals might allow for re-scaling experimental errors
by a Birge factor, something that, to our knowledge, remains to be
established. We do not expect, however, this to be the case. The
Granada analysis including all $\sim 8000$ NN data is similar in size
and provides a similar reduced $\chi^2/N \sim 1.7$. Only after
self-consistent $3\sigma$ data selection down to $\sim 6700$ data is
the normality test passed. The database can be downloaded from
Ref.~\refcite{GranadaDB}.

Likewise, the widely used Optimized Chiral Nucleon-Nucleon Interaction
at Next-to-Next-to-Leading Order \cite{Ekstrom:2013kea} fits $N=1945$
scattering NN data for laboratory energies below 125 MeV, yielding a
reduced $\chi^2/{\rm d.o.f.} = 1.15$, already $5$ $\sigma$ excluded as a
consistent fit. Unfortunately, as we have recently
shown~\cite{Perez:2014bua}, their residuals do not pass the normality
test, so that one cannot re-scale by a Birge factor the experimental
uncertainties, indicating that the fit is inconsistent. Therefore,
either some data are inconsistent among themselves or the proposed
theory contains sizable systematic errors.

\subsection{Fitting scattering data vs fitting phase-shifts}

However, carrying out a large scale PWA implementing all necessary
physical effects is rather messy, and one may naturally wonder to what
extent all these complications are necessary. Moreover, the main
outcome of the PWA are the phase-shifts and mixing parameters. Thus,
quite often researchers prefer to fit phase shits from any of the
existing data analyses.  The key question is: when do the phases
extracted from a PWA qualify for tracking the experimental NN
uncertainties?. Firstly, one must have for the reduced $\chi^2$ a
given confidence interval, i.e. $\chi^2/{\rm d.o.f.} = 1 \pm
\sqrt{2/{\rm d.o.f.}}$ in order to guarantee a sensible fit with a
statistically meaningful confidence level. This is certainly a
sufficient condition to carry out a statistically error analysis. Does
this mean that if the reduced $\chi^2$ does not fall within this range
we have a bad fit?.  Not necessarily. Fortunately, the reduced
$\chi^2$ condition is not a necessary one. If the value falls outside
this interval, one may still re-scale the errors by a so-called Birge
factor, {\it provided} one can test positively the hypothesis that the
set of residuals are a scaled normal variable.  Thus, this requires
passing a normality test for the residuals. We have repeatedly
stressed this aspect~\cite{Perez:2014yla,Perez:2014kpa} to justify the
selection of our database in Ref.~\refcite{Perez:2013jpa} and to
promote as high quality the 6 interactions proposed by
us~\cite{Perez:2014waa} and used here in our study of Skyrme
parameters and counter-terms.

Assuming that all these conditions are fulfilled one would obtain from
the PWA a set of phase-shifts with legitimate statistical error bars
stemming from experimental uncertainties. Unfortunately, this is a
simplifying assumption for it ignores existing correlations among the
set of partial waves induced by the fit of scattering observables
directly to the available scattering data. Our experience on replacing
the full PWA by a reduced set of phase-shifts and mixing parameters is
not positive as far as the complete $\chi^2/{\rm d.o.f.}$ is
concerned.  Moreover, this shortcut does not prove to be a faithful
determination of statistical errors. This is certainly a regrettable
situation which finds its origin, as outlined above, on the lack of
control of predicting non-fitted observables, see the discussion
around Fig.~\ref{fig:DtObservable}, ultimately triggered by the
abundance bias as well as the specific choice of interaction. Under
these circumstances we recommended~\cite{Perez:2014waa} that if a fit
to phase-shifts was to be carried out, probably the least biased
strategy would be fitting to an average and standard deviation of the
6 proposed potentials, which provides also a larger error bar and for
which complete tables are provided. As said, this is based on our own
experience on performing fits to single and separate phase-shifts
where only statistical error bars were included; the inferred
interaction did not reproduce accurately the scattering data from
which the phase-shifts being fitted were deduced.

In what follows we comment on recent work where the phase-shift
strategy was pursued and also on the possible pitfalls. The local
chiral potential suggested in Ref.~\refcite{Gezerlis:2014zia} and
based on phase-shift fit has been tested in our recent work against
the corresponding experimental data within the same energy
range~\cite{Perez:2014bua} providing extremely large reduced $\chi^2
/N \sim 10-10^4$ depending on the maximal energy.

While the calculation in Ref.~\refcite{Epelbaum:2014efa} makes, in the
authors words, obsolete the widely advertised previous versions, one
should say that in line with all their previous developments they have
not been confronted with scattering data directly but rather to phase
shifts obtained in the SAID analysis~\cite{SAID}. While abundant
information regarding good visual fits against selected data is
provided, no $\chi^2$ to the total number of data used to extract the
phase-shifts has been reported. As we pointed out before the, SAID
$\chi^2/N$ gives $1.7$ for the analyzed data bellow laboratory energy of
$350$ MeV, which is $25$ $\sigma$ away from any reasonable confidence
level.

Recently a benchmarking analysis of peripheral nucleon-nucleon
scattering at fifth order of chiral perturbation theory using input
from $\pi N$ scattering was carried out~\cite{Entem:2014msa}. There a
comparison to peripheral phase shifts was shown and good visual
agreement can be seen, with the single exception of the $^1F_3$ wave,
when compared to the Nijmegen and SAID databases~\cite{SAID} is
undertaken. However, as it is well known, peripheral phase-shifts can
only be extracted from experiment after a complete PWA, thus they are
only indirectly accessible. Besides this, peripheral waves are
determined very accurately, as they mostly stem from the OPE potential
tail and their error bars are actually tiny. As our recent
analysis~\cite{Perez:2014waa} shows, uncertainties go beyond what a
visual fit may discern. Unfortunately, the analysis of
Ref.~\refcite{Entem:2014msa} does not provide a quantitative measure
of the agreement. This is actually a place where a scrupulous error
analysis might provide extremely useful information on the validation
of the theory.

\subsection{Bayesian vs Classical}

The Bayesian framework is a particularly appealing approach as it
mainly poses the question of the chances for the theory  being
correct given some data rather than the chances for the data  being
correct under the assumption that the testing theory is the right
one. In the Bayesian interpretation, the fitting parameters become
random variables which are determined from the given given data and
assuming a prior probability of finding the parameters independently
on the experimental data subjected to the analysis. Besides the purely
ontological aspects, the Bayesian approach is practical as it educates
the theory when the number of parameters and the number of data are
similar. A further rewarding aspect of the approach is the asymptotic
consistency between both Classical and Bayesian approaches when the
number of data actually becomes much larger than the number of
parameters.

The practical implementation proceeds via the augmented
$\chi^2$-approach where an additional term is added to the standard
$\chi^2$, see Eq.~\ref{eq:chi2-exp}, and which will be denoted as
$\chi^2_{\rm exp}$ here. This new term will be denoted as $\chi^2_{\rm
  th}$ incorporating some { \it a priori} known and fuzzy constraints
on the fitting parameters and are incorporated as follows (see
e.g. Refs.~\refcite{Lepage:2001ym,Morningstar:2001je,Schindler:2008fh}
for pedagogical introductions)
\begin{equation}
\chi_{\rm th}^2 =  \sum_{i=1}^P \left(\frac{p_i -
  p_i^{\rm th}}{\Delta p_i} \right)^2,
\end{equation}
where one {\it expects} the parameters $p_i$ to be within $p_i^{\rm
  th} \pm \Delta p_i$. The case of absolute ignorance corresponds to
$\Delta p_i \to \infty $ where the case of absolute certainty
corresponds to $\Delta p_i \to 0$. Let us denote by $N_p$ the number
of effective constraints, i.e. the number of parameters where $0 <
\Delta p_i < \infty$. As mentioned, the impact of this new term can
only be sizable under certain operating conditions. Namely, $\chi_{\rm
  exp}^2 \sim \chi^2_{\rm th}$. Otherwise, some {\it ad hoc} choice on
the relative weight of both contributions has to be made. This is a
key question and depends both on the quality of the data as well as
the confidence of the constraints. Obviously, for a small number of
constraints $N_p$ as compared to the number of data or pseudo-data
$N_d$, a direct addition of $\chi^2_{\rm exp}$ and $\chi^2_{\rm th}$
would make the constraints irrelevant. Therefore, and following a
suggestion~\cite{Ledwig:2014cla}
\footnote{The method has been applied for the case of meson-meson
  scattering~\cite{Ledwig:2014cla} where the fitted data were not
  provided with any error bar estimates and some of the fitting
  parameters had natural $1/N_c$ accurate estimates.}  one may
construct a {\it reduced} $\chi^2$, $\bar \chi^2 \equiv \chi^2 /N $,
with a $50\%$ weighting on the data/pseudo-data and the theoretical
constraints, corresponding to 
\begin{eqnarray}
  \chi^2 _{\rm total}\; =\; \frac{N_d+N_p}2\; \left( \frac{\chi^2_{\rm
      exp}}{N_{d}} + \frac{\chi^2_{\rm th}}{N_p} \right)\, .
  \label{eq:chi2tot}
\end{eqnarray}
The additional terms in the total $\chi^2$ impose a penalty for fits
which deviate from the $N_p$ theoretical expectations on the fitting
parameters $p_i$ significantly from the {\it a priori} theoretical
expectation $p_i^{\rm th} \pm \Delta p_i$. In the case under study
corresponding to NN scattering our fitting parameters are the
strengths of delta-shells or Gaussian radial functions for which no
obvious guess is available. In the present study the counter-terms are
taking as derived secondary quantities, but it would certainly be very
interesting to test their size in a Bayesian manner based on
theoretical expectations deduced either from chiral perturbation
theory or large $N_c$ arguments as successfully done in a recent
meson-meson scattering analysis~\cite{Ledwig:2014cla}.

Actually, a similar idea was implemented
recently~\cite{Epelbaum:2014efa} to impose the expected $SU(4)$ Wigner
symmetry for S-waves, $\tilde C_{^1S_0} \sim \tilde C_{^3S_1} $, in a
Bayesian fashion using the augmented $\chi^2$ realization. As we see
from Table \ref{tab:Cterms} even when we take into account our
estimate for the systematic error these two numbers differ. 

Of course, one possible interpretation of the Wigner symmetry as a
long distance symmetry is Wilsonian in nature as suggested in
Ref.~\refcite{Arriola:2010hj} and checked within the Similarity
Renormalization group
approach~\cite{Timoteo:2011tt,Arriola:2013nja,Arriola:2013era,Arriola:2014fqa}
at relatively small energy scales. In such a case the violation
reported in table \ref{tab:Cterms} based on a higher energies analysis
is not significant as far as the Wigner symmetry analysis is
concerned, and may introduce a bias in the Bayesian constraint invoked
in Ref. \refcite{Epelbaum:2014efa}.

\section{Conclusions}

We have extracted the Skyrme parameters of effective interactions from
six phenomenological realistic NN interactions. For each potential the
statistical uncertainty from the experimental NN scattering data was
propagated into the calculated Skyrme parameters. The statistical
uncertainties of the parameters are the same order of magnitude for
the six potentials. In most cases the extracted Skyrme parameters from
different interactions are incompatible within 1 standard
deviation. Since the six interactions are statistically equivalent,
their discrepancies must originate from the different representations
of the NN interaction and are therefore considered to be systematic
uncertainties. We estimated the systematic uncertainty of the Skyrme
parameters with the sample standard deviation of the six interactions
and found it to be at least an order of magnitude larger than the
statistical ones. The same procedure was followed to estimate the
systematic uncertainties of the counter-terms, which resulted to be
also an order of magnitude larger than the statistical ones.

We have also discussed and shown how a detailed statistical scrutiny
of the NN scattering data may provide valuable hints on the interplay
between theory and experiment and their assumed
uncertainties. Ignoring these important pieces of information is not
only misleading, but enhances strongly biased views on the underlying
dynamical structure of the nuclear force and promoting at the same
time interactions which are not properly validated against the
available data. These are crucial points for the predictive power of
theoretical nuclear physics, since the reliable uncertainty
quantification of nuclear forces is an urgent necessity in {\it ab
  initio} nuclear structure and nuclear reactions calculations.

\section*{Acknowledgements}
This work is supported by Spanish DGI (grant FIS2014-59386-P) and
Junta de Andaluc{\'{\i}a} (grant FQM225). This work was partly
performed under the auspices of the U.S. Department of Energy by
Lawrence Livermore National Laboratory under Contract
No. DE-AC52-07NA27344.  Funding was also provided by the U.S.
Department of Energy, Office of Science, Office of Nuclear Physics
under Award No.  DE-SC0008511 (NUCLEI SciDAC Collaboration).

\bibliographystyle{ws-ijmpe} \bibliography{references}
 
\end{document}